\documentclass[aps,showpacs,preprint,superscriptaddress]{revtex4}
\usepackage{graphicx}
\usepackage{amsmath}

\begin{document}

\title{Vacuum pair production of charged scalar bosons in time-dependent electric fields}

\author{Zi-Liang Li}
\author{Ding Lu}
\author{Bai-Song Xie\footnote{Corresponding author. Email address: bsxie@bnu.edu.cn}}
\affiliation{Key Laboratory of Beam Technology and Materials Modification of the Ministry of Education,
College of Nuclear Science and Technology, Beijing Normal University, Beijing 100875, China}

\begin{abstract}
Based on the quantum mechanical scattering model, the dynamical assist effect and the multiple-slit interference effect in electron-positron pair production from vacuum are generalized to vacuum pair production of charged scalar bosons. For the former effect some combinations of a strong but slowly varying electric field and a weak but rapidly varying one with different time delay are studied. Results indicate that the oscillation intensity of momentum spectrum and the number density of created bosons reduce with increasing of the time delay. Obviously, they achieve the maximum if the time delay equals zero. For the latter effect, it is shown that this effect does not exist for equal-sign $N$-pulse electric field in contrast to its existence for alternating-sign $N$-pulse. An approximate solution of boson momentum spectrum is got and it is agreeable well with the exact numerical one in alternating-sign $N$-pulse electric field, especially for $2$-pulse field and for small longitudinal momentum. The difference of vacuum pair production between bosons and fermions are also compared for their longitudinal momentum spectra.

\emph{\textbf{keywords:}} Vacuum pair production; Dynamical assist effect; Multiple-slit interference effect.

\end{abstract}
\pacs{12.20.Ds, 11.15.Kc, 03.65.Sq}

\maketitle

\section{Introduction}

One of the most astonishing theoretical predictions of quantum electrodynamic (QED) is a finding that the vacuum in a strong external field is unstable and will decay into electron-positron pairs. This process as one of the most interesting phenomena in strong field physics had been studied several decades ago. After the earlier work of Sauter \cite{Sauter} and Heisenberg \cite{Heisenberg}, Schwinger \cite{Schwinger} first got the vacuum pair production rate in constant electric field by calculating the imaginary part of one loop effective lagrangian in the middle of last century and obtained the critical electric field strength  $E_{cr}=m_e^2c^3/e\hbar\sim1.32\times10^{16}\mathrm{V}/\mathrm{cm}$, where $-e$ and $m_e$ are the charge and rest mass of electrons, $c$ is the speed of light in vacuum, and $\hbar$ is the Planck constant. This critical electric field strength corresponds to the laser intensity of $\sim4\times10^{29}\mathrm{W}/\mathrm{cm}^2$. Now the electron-positron pair production from vacuum is also called Schwinger mechanism.

Since Schwinger's pioneering study many works have been done including the new research methods and the different forms of external fields. For example, the employed methods include the semiclassical one, such as WKB approximation \cite{Kim} and worldline instanton technique \cite{Dunne}, and the quantum kinetic method, such as quantum Vlasov equation (QVE) \cite{Kluger,Schmidt,Alkofer,Roberts} and Wigner formalism \cite{Hebenstreit}. By using these methods, Schwinger mechanism for spatially homogenous but time-dependent electric fields and inhomogenous static electric fields with or without a constant magnetic field have been investigated. Although many theoretical works are performed, however, the experimental observation of this mechanism is still a challenging task.

Recent years, as the rapid development of laser technology, hopes for realizing Schwinger mechanism in laboratory are rekindled. Now the ultrahigh intensity laser facilities which are building such as the Extreme Light Infrastructure (ELI) \cite{ELI} and the European X-ray free electron laser (XFEL) \cite{Ringwald} are possible to approach the subcritical field strength, i.e., $E\sim0.1E_{cr}$. Nevertheless, due to the exponentially suppression of Schwinger mechanism, which can be regarded as tunneling effect, the number of produced pairs in a subcritical field is very small. To improve the pair creation number, some catalytic mechanisms are suggested in the future experiment \cite{Schutzhold,Piazza,GVDunne,Bulanov}. Among them, the dynamical assist Schwinger mechanism draws an attention \cite{Schutzhold,Orthabera,Fey}. In subcritical field regime, by superimposing a weak strength but rapidly varying electric field (dominated by multiphoton mechanism) on a strong strength but slowly varying one (dominated by Schwinger mechanism), the number of produced particles can be enhanced greatly, while both of these two fields have a suppressed pair creation number when they are considered respectively. Moreover, this effect is also generalized to more realistic laser pulse with the optimization of electric fields \cite{Nuriman1,Kohlfurst,Nuriman2}. In addition, in Ref. \cite{Akkermans}, the time-dependent multiple-slit interference effect for Schwinger mechanism is studied. It is found that the maxima of longitudinal momentum spectrum can reach $N^2$ times the single pulse profile for alternating-sign $N$-pulse electric field because its coherent enhancement effect. It seems a good candidate to realize the Schwinger mechanism in laboratory.

The propose of this work is to generalize the dynamical assist effect and the multiple-slit interference effect in electron-positron pair production to charged scalar boson pair production. For the former effect, we study the role of a time delay playing on the boson pair production in two kinds of combinational electric fields, i.e., equal-sign electric field and alternating-sign electric field. The combination fields constitute by a strong but slowly varying electric field and a weak but rapidly varying one with different time delays. For the latter effect, we investigate it in equal-sign and alternating-sign electric field pulse train. We give the exact turning points of alternating-sign $2$-pulse electric field and use them to calculate the approximate solution of momentum spectrum in both $2$-pulse and $N$-pulse electric field. A simple expression of approximate solution in alternating-sign $N$-pulse electric field for boson pair production is also got. For both of two effects, the different physical properties of bosons and fermions are shown by comparing their longitudinal momentum spectra of created particles.

The paper is organized as following. In Sec. II, from Klein-Gordon equation we simply deduce a Riccati equation and give an approximate solution of momentum distribution function for boson pair production in the frame of quantum mechanical scattering model. In Sec. III  we study the dynamical assist effect in two different combinational electric fields, and the multiple-slit interference effect in $2$-pulse and $N$-pulse electric field. The longitudinal momentum spectra of bosons and fermions are compared with each other. Section IV gives the discussions and conclusions for our results.

\section{Theoretical analysis by the scattering model}

Here we consider a standing-wave electric field produced by two counter propagating laser pulses. Since the spatial focusing scale of laser pulses is larger than our studied region, i.e., Compton wavelength, the spatial variations of external fields are neglected. Then the magnetic effects are ignored, and we get a spatially homogenous and time-dependent electric field $\mathbf{E}(t)=(0,0,E(t))$ that is along the $x^3$ axis. By using the temporal gauge $A^0(t)=0$, the vector potential is given as $A^\mu(t)=(0,0,0,A(t))$, with $E(t)=-\dot{A}(t)$. For convenience, natural units $\hbar=c=1$ are chosen.

\subsection{Riccati equation}

From Klein-Gordon equation $(D_\mu D^\mu+m^2)\Psi(\mathbf{x},t)=0$, where $D_\mu=\partial_\mu+iqA_\mu$, and $q$ and $m$ are the charge and rest mass of bosons, we can get
\begin{equation}
\big[\partial_t^2-\nabla^2+2iqA(t)\partial_3+q^2A^2(t)+m^2\big]\Psi(\mathbf{x},t)=0.\label{eq1}
\end{equation}
The space-independent vector potential makes it convenient to have a Fourier mode decomposition $\Psi(\mathbf{x},t)=\int\frac{d^3k}{(2\pi)^3}e^{i\mathbf{k}\cdot\mathbf{x}}\psi(\mathbf{k},t)$, and the Fourier modes $\psi(\mathbf{k},t)$ satisfy the Schr\"{o}dinger-like equation:
\begin{equation}
\frac{d^2\psi(\mathbf{k},t)}{dt^2}+\Omega^2(\mathbf{k},t)\psi(\mathbf{k},t)=0,\label{eq2}
\end{equation}
where
\begin{equation}
\Omega(\mathbf{k},t)=\sqrt{m^2+\mathbf{k}_\perp^2+[k_\parallel-qA(t)]^2} \label{eq3}
\end{equation}
depends on the $k_\parallel$ and $\mathbf{k}_\perp$, the parallel and perpendicular component of canonical momentum with respect to the electric field direction.

Following the routine of Ref.\cite{Popov}, we assume
\begin{eqnarray}
&\psi(\mathbf{k},t)=Q(\mathbf{k},t)\Big[e^{-i\Theta(\mathbf{k},t)}
+R(\mathbf{k},t)e^{i\Theta(\mathbf{k},t)}\Big],&\label{eq4}\\
&\dot{\psi}(\mathbf{k},t)=i\Omega(\mathbf{k},t)Q(\mathbf{k},t)\Big[-e^{-i\Theta(\mathbf{k},t)}
+R(\mathbf{k},t)e^{i\Theta(\mathbf{k},t)}\Big],&\label{eq5}
\end{eqnarray}
where
\begin{equation}
\Theta(\mathbf{k},t)=\int_{-\infty}^t\Omega(\mathbf{k},t^\prime)dt^\prime \label{eq6}
\end{equation}
is the phase factor in Fourier $\mathbf{k}$ space.
From Eq. (\ref{eq2}) and Eqs. (\ref{eq4}-\ref{eq5}), it is easy to find that $R(\mathbf{k},t)$ satisfies the Riccati equation
\begin{equation}
\dot{R}(\mathbf{k},t)=\frac{\dot{\Omega}(\mathbf{k},t)}{2\Omega(\mathbf{k},t)}\Big[e^{-2i\Theta(\mathbf{k},t)}
-R^2(\mathbf{k},t)e^{2i\Theta(\mathbf{k},t)}\Big],\label{eq7}
\end{equation}
where $R(\mathbf{k},-\infty)=0$ is the initial condition and $|R(\mathbf{k},+\infty)|^2$ is exactly the reflection coefficient in the view of quantum mechanical scattering theory. Obviously, $|R(\mathbf{k},+\infty)|^2$ can be got by numerically solving Eq. (\ref{eq7}) with the given initial condition. On the basis of  Nikishov's calculation \cite{Nikishov}, the momentum distribution function of created scalar bosons is defined as
\begin{equation}
f(\mathbf{k})\equiv\frac{|R(\mathbf{k},+\infty)|^2}{1-|R(\mathbf{k},+\infty)|^2}.\label{eq8}
\end{equation}
Since vacuum pair production is exponentially suppressed with respect to external electric field strength, $|R(\mathbf{k},+\infty)|^2$ is much less than $1$ in subcritical field strength. Eq. (\ref{eq8}) is reduced to $f(\mathbf{k})\approx|R(\mathbf{k},+\infty)|^2$. In addition, the number density of produced boson pairs can be conveniently got by the definition as
\begin{equation}
n(+\infty)\equiv\int \frac{d^3k}{(2\pi)^3}f(\mathbf{k}).\label{eq9}
\end{equation}

\subsection{Approximate solution}

It is well known that $R(\mathbf{k},t)$ is very small in subcritical electric field. So $R(\mathbf{k},+\infty)$ of Eq.(\ref{eq7}) can be approximately calculated as
\begin{equation}
R(\mathbf{k},+\infty)\approx\int_{-\infty}^\infty\frac{\dot{\Omega}(\mathbf{k},t)}{2\Omega(\mathbf{k},t)}
e^{-2i\Theta(\mathbf{k},t)}.\label{eq10}
\end{equation}
It can be seen that the main contributions to the above integral are from the vicinity of turning points $t_P$ which satisfy $\Omega(\mathbf{k},t_P)=0$. Furthermore, assuming that all of the turning points are fully separated, the approximate solution of $R(\mathbf{k},+\infty)$ becomes (refer to Refs. \cite{Meyer,Dumlu1})
\begin{equation}
R(\mathbf{k},+\infty)\approx\sum_{t_P}e^{-2i\int_{-\infty}^{t_P}\Omega(\mathbf{k},t)dt}, \label{eq11}
\end{equation}
where $t_P$ denotes the turning points in the upper half complex $t$ plane. Then the momentum distribution function of the created bosons is
\begin{equation}\label{eq12}
f(\mathbf{k})\approx\sum_{t_P}e^{-2\vartheta^P(\mathbf{k})}+\sum_{t_P\neq t_{P^\prime}}
2\cos[2\theta^{(P,P^\prime)}(\mathbf{k})]e^{-\vartheta^P(\mathbf{k})-\vartheta^{P^\prime}(\mathbf{k})},
\end{equation}
where $\vartheta^P(\mathbf{k})=\Big|\int_{t_P^*}^{t_P}\Omega(\mathbf{k},t)dt\Big|$, $\theta^{(P,P^\prime)}(\mathbf{k})=\int_{\mathrm{Re}(t_P)}^{\mathrm{Re}(t_{P^\prime})}\Omega(\mathbf{k},t)dt$, and the prime denotes different pairs of turning points. Obviously the dominant contributions to pair production come from the turning points which are closest to the real axis of complex $t$ plane. Another fact is that the interference between different pairs of turning points depends on the distance between them along the real axis of complex $t$ plane.

\section{Results}

In this section, the dynamical assist effect and the multiple-slit interference effect of charged scalar boson pair production are investigated. The results are also compared with fermion pair production. For convenience, we set $q=m=1$ and $\mathbf{k}_\perp=0$.

\subsection{Dynamical assist effect}

In Ref. \cite{Schutzhold}, it presents that the superposition of a strong but slowly varying electric field with a weak but rapidly varying electric field significantly enhances the vacuum electron-positron pair creation. This can be intuitively interpreted as the dynamical assist effect of perturbative multi-photon mechanism to non-perturbative Schwinger mechanism. To study this effect in boson pair production, we consider some combinations of a strong but slowly varying electric field $E_1(t)$ and a weak but rapidly varying one $E_2(t)$ with different time delays.

\emph{1. Equal-sign electric field}

The first kind of combinational electric field is described by
\begin{eqnarray}
E_{A1}(t)&=&E_1(t)+E_2(t) \nonumber\\
&=&E_1\mathrm{sech}^2\Big[\omega_1 \Big(t+\frac{T}{2}\Big)\Big]+E_2\mathrm{sech}^2\Big[\omega_2 \Big(t-\frac{T}{2}\Big)\Big], \label{eq13}
\end{eqnarray}
where $E_1$ and $E_2$ are the amplitudes of electric field, $\omega_1$ and $\omega_2$ are the inverse pulse widths of field, and $T$ denotes the time delay between $E_1(t)$ and $E_2(t)$. The electric field parameters are chosen as $E_1=0.25$, $\omega_1=0.02$, and $E_2=0.025$, $\omega_2=1$.

\begin{figure}[tbp]\suppressfloats
\centering
\includegraphics[width=10cm]{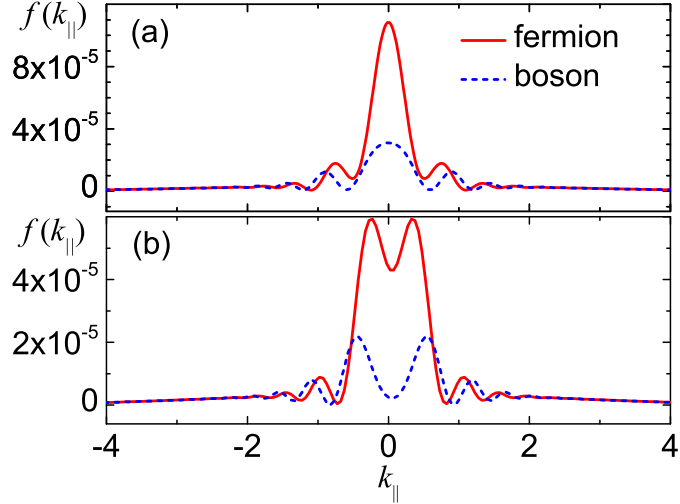}
\caption{\label{Fig1}(color online) The longitudinal momentum spectra of created bosons (dashed blue line) and fermions (solid red line) in, (a) equal-sign electric field $E_{A1}(t)$, and (b) alternating-sign electric field $E_{A2}(t)$. The electric field parameters are $E_1=0.25$, $\omega_1=0.02$, and $E_2=0.025$, $\omega_2=1$ with the time delay $T=0$.}
\end{figure}

When $T=0$, the combinational electric field $E_{A1}$ is symmetric and has a maximum value $E_1+E_2$ at $t=0$. In this case, by numerically solving Eqs. (\ref{eq7}) and (\ref{eq8}), the longitudinal momentum spectrum of created bosons (dashed blue line) is obtained and shown in Fig. \ref{Fig1}(a). One can see clearly that the oscillatory structure of momentum spectrum is distinct. And the dynamical assist effect is still valid to boson pair creation, see Fig. \ref{Fig3}.

\begin{figure}[tbp]\suppressfloats
\centering
\includegraphics[width=10cm]{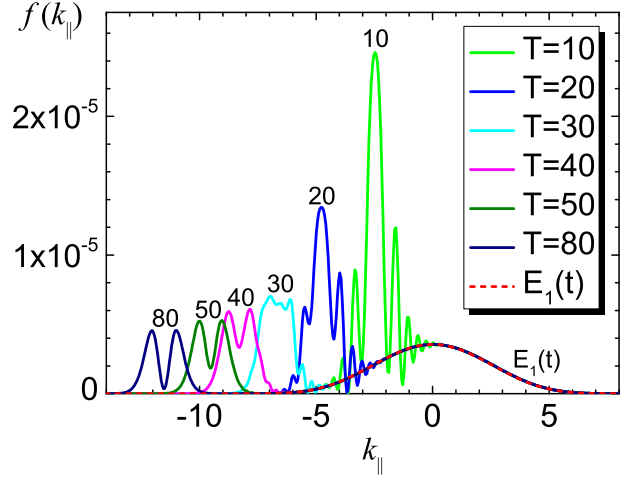}
\caption{\label{Fig2}(color online) The longitudinal momentum spectrum of created bosons changing with the time delay $T$ in equal-sign electric field $E_{A1}(t)$. The electric field parameters are $E_1=0.25$, $\omega_1=0.02$, and $E_2=0.025$, $\omega_2=1$.}
\end{figure}

When $T\neq0$, the longitudinal momentum spectrum of created bosons changing with the time delay $T$ are presented in Fig. \ref{Fig2}. The values of $T$ are chosen as $T=10$ (green line), $20$ (blue line), $30$ (cyan line), $40$ (Magenta line), $50$ (Olive line) and $80$ (Navy line). For a reference, the momentum spectrum of created bosons in the strong but slowly varying electric field $E_1(t)$ (red dashed line) are also plotted. It is easy to find that the oscillation behaviors become weaker and weaker as the increase of time delay. Particularly, when $T\gtrsim80$, the oscillations are almost disappeared and the momentum distribution in the electric field $E_1(t)$ and $E_2(t)$ are well separated. Meanwhile, the dynamical assist effect is not obvious any more, see Fig. \ref{Fig3}. With the increase of $|T|$, the number density created in the combinational electric field $E_{A1}(t)$ becomes smaller and smaller and finally equals to the sum of the particle number density in the electric field $E_1(t)$ and $E_2(t)$. Thus, to increase the probability of pair production, it is necessary to shorten the time delay between two fields. Note that when the longitudinal momentum $k_\parallel>0$, the momentum spectrum in the combinational electric field $E_{A1}(t)$ and the electric field $E_1(t)$ are almost overlapping. However, there are still some small differences between them because the created bosons in the electric field $E_1(t)$ will be accelerated by $E_2(t)$. This result is similar to that of Ref. \cite{Orthabera}.

\begin{figure}[tbp]\suppressfloats
\centering
\includegraphics[width=10cm]{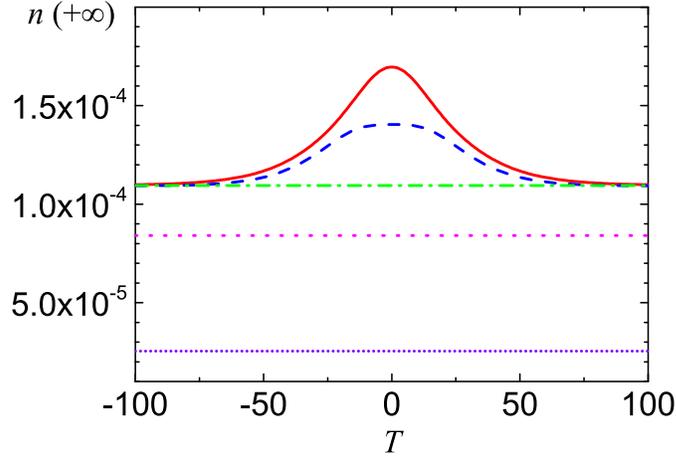}
\caption{\label{Fig3}(color online) The number density of created boson pairs $n(+\infty)$ changing with the time delay $T$ in equal-sign electric field $E_{A1}(t)$ (solid red line), alternating-sign electric field $E_{A2}(t)$ (dashed blue line), the strong but slowly varying electric field $E_1(t)$ (dotted magenta line), and the weak but rapidly varying electric field $\pm E_2(t)$ (short dotted violet line). The dashed-dotted green line denotes the sum of the produced particle number density in electric field $E_1(t)$ and $\pm E_2(t)$. The electric field parameters are $E_1=0.25$, $\omega_1=0.02$, and $E_2=0.025$, $\omega_2=1$.}
\end{figure}

\emph{2. Alternating-sign electric field}

Now we consider the second kind of combinational electric field constituted by the strong but slowly varying electric field $E_1(t)$ minus the weak but rapidly varying one $E_2(t)$,
\begin{eqnarray}
E_{A2}(t)&=&E_1(t)-E_2(t) \nonumber\\
&=&E_1\mathrm{sech}^2\Big[\omega_1 \Big(t+\frac{T}{2}\Big)\Big]-E_2\mathrm{sech}^2\Big[\omega_2 \Big(t-\frac{T}{2}\Big)\Big]. \label{eq14}
\end{eqnarray}
The parameters are chosen the same as the former case as $E_1=0.25$, $\omega_1=0.02$, and $E_2=0.025$, $\omega_2=1$.

\begin{figure}[tbp]\suppressfloats
\centering
\includegraphics[width=10cm]{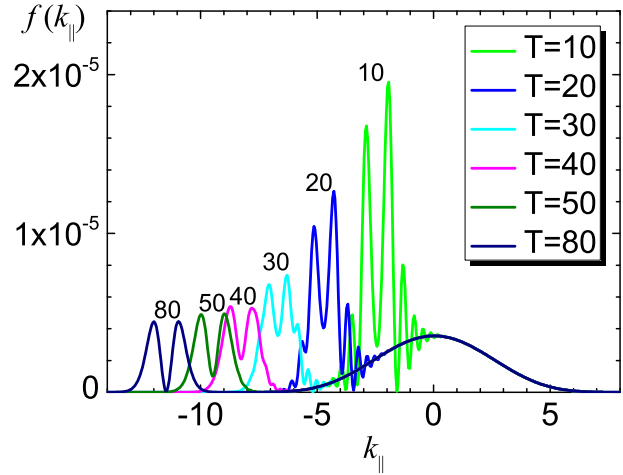}
\caption{\label{Fig4}(color online) The longitudinal momentum spectrum of created bosons changing with the time delay $T$ in alternating-sign electric field $E_{A2}(t)$. The electric field parameters are $E_1=0.25$, $\omega_1=0.02$, and $E_2=0.025$, $\omega_2=1$.}
\end{figure}

When $T=0$, the combinational electric field $E_{A2}(t)$ is also symmetric and has a local minimal value $E_1-E_2$ at $t=0$. In this situation, the longitudinal momentum spectrum of created bosons (dashed blue line) is calculated and plotted in Fig. \ref{Fig1}(b). We can see that there are two noticeable differences from the momentum spectrum in Fig. \ref{Fig1}(a). First, the momentum distribution function $f(k_\parallel)$ is smaller than that in the equal-sign combinational electric field $E_{A1}(t)$ because the strong but slowly varying electric field $E_1(t)$ is weakened by the weak but rapidly varying one $E_2(t)$. Second, for small longitudinal momentum $k_\parallel$, the single hump structure of the momentum spectrum in Fig. \ref{Fig1}(a) is replaced by a double hump one. Moreover, the dynamical assist effect still exists in the alternating-sign combinational electric field $E_{A2}(t)$ (dashed blue line), see Fig. \ref{Fig3}.

When the time delay $T\neq0$, the longitudinal momentum spectrum of created bosons changing with $T$ are shown in Fig. \ref{Fig4}. One can see that its variation trend is similar to that in the equal-sign combinational electric field $E_{A1}(t)$. Nevertheless, the dynamical assist effect is always weaker than that in the plus field $E_{A1}(t)$ and there is a flat top region of the created pair number density $n(+\infty)$. It indicates that in this region the dynamical assist effect is insensitive to the time delay, while beyond it the particle number density will rapidly decrease as the increase of $|T|$. Note that when $|T|\rightarrow100$, the dynamical assist effect will completely vanish because the complete separation occurs between the electric field $E_1(t)$ and $E_2(t)$.

\subsection{Multiple-slit interference effect}

In Ref. \cite{Akkermans}, the time-domain multiple-slit interference effect for electron-positron pair production in the sequences of alternating-sign electric field pulses is investigated. Here we shall generalize it to the pair production of charged scalar boson.

\emph{1. 2-pulse electric field}

We begin the study to the 2-pulse electric field composed by two Sauter-like electric fields as
\begin{equation}
E_{B1\pm}(t)=E_0\mathrm{sech}^2\Big[\omega_0 \Big(t+\frac{T}{2}\Big)\Big]\pm E_0\mathrm{sech}^2\Big[\omega_0 \Big(t-\frac{T}{2}\Big)\Big], \label{eq15}
\end{equation}
where $E_0$ is the amplitude of electric field, $\omega_0$ represents the inverse width scale, $T$ is the time delay, ``+'' denotes equal-sign electric field pulse trains, and ``-'' is alternating-sign electric field pulse trains. The parameters of field are chosen as $E_0=0.1$ and $\omega_0=0.05$.

\begin{figure}[tbp]\suppressfloats
\centering
\includegraphics[width=10cm]{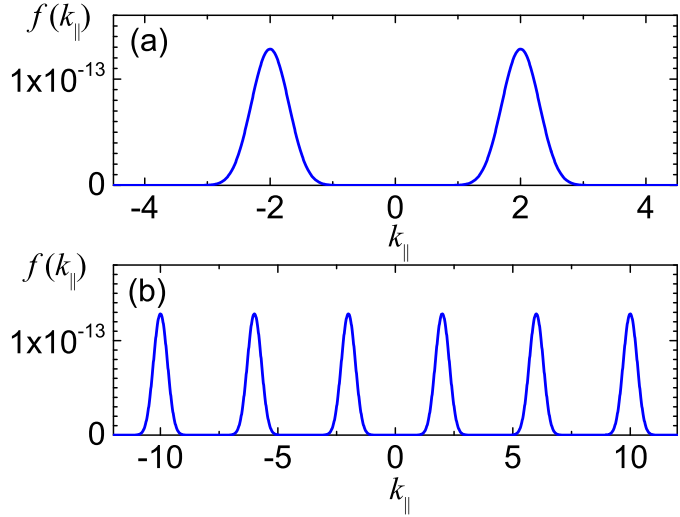}
\caption{\label{Fig5}(color online) The longitudinal momentum spectrum of created bosons in, (a) equal-sign $2$-pulse electric field $E_{B1+}(t)$, and (b) equal-sign $N(N=6)$-pulse electric field $E_{B2+}$. The electric field parameters are $E_0=0.1$, $\omega_0=0.05$, and $T=180.32$.}
\end{figure}

For the equal-sign electric field pulse trains $E_{B1+}(t)$, there is only one pair of turning points, which plays the leading role in pair creation. Therefore, no interference effect presents in the longitudinal momentum spectrum of created bosons, see Fig. \ref{Fig5}(a). Moreover, the momentum spectrum in the first pulse electric field separates from that in the second electric field because the created pairs in the first electric field are accelerated by the second one.

\begin{figure}[tbp]\suppressfloats
\centering
\includegraphics[width=10cm]{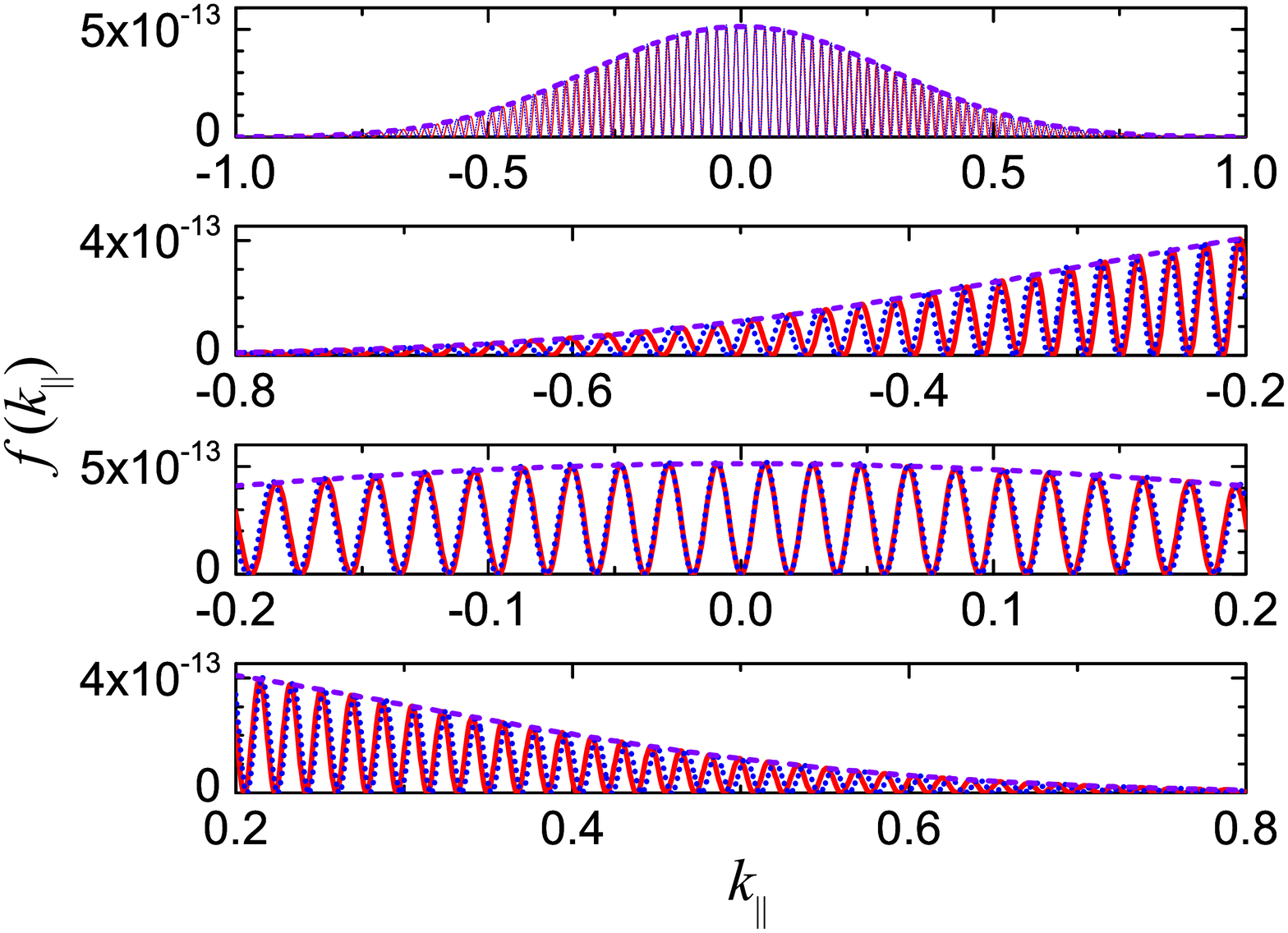}
\caption{\label{Fig6}(color online) The longitudinal momentum spectrum of created bosons in alternating-sign $2$-pulse electric field $E_{B1-}(t)$. The solid red line is the numerical result, and the dotted blue line is the approximate solution. The lower three panels are the local amplification of the top one. The electric field parameters are $E_0=0.1$, $\omega_0=0.05$, and $T=180.32$.}
\end{figure}

For the alternating-sign electric field pulse trains $E_{B1-}(t)$, the vector potential is $A(t)=E_0/\omega_0\{1+\tanh[\omega_0(t-T/2)]-\tanh[\omega_0(t+T/2)]\}$. By solving the algebraic equation $\Omega(\mathbf{k},t)=0$, the exact turning points in the upper half complex $t$ plane can be conveniently got as
\begin{equation}
t_\pm=\pm\frac{1}{\omega_0}\mathrm{arctanh}\sqrt{\frac{\omega_0(k_\parallel\pm i)
+2E_0\tanh(\frac{\omega_0T}{2})}{[\omega_0(k_\parallel\pm i)-E_0]\tanh^2(\frac{\omega_0T}{2})
+2E_0\tanh(\frac{\omega_0T}{2})}}+\frac{ip\pi}{\omega_0}, \label{eq16}
\end{equation}
where $p=0,1,2,...$ are the nonnegative integer. Since the dominant contributions to $f(\mathbf{k})$ come from the turning points closest to the real axis of complex $t$ plane, it is a good approximation to set $p=0$, and only two complex conjugate pairs of turning points, which have the same distance from the real axis in $t$ plane, decide the momentum spectrum. This simple analysis means that $\vartheta^+(\mathbf{k})=\vartheta^-(\mathbf{k})\equiv\vartheta(\mathbf{k})$, where $\vartheta^+(\mathbf{k})=\Big|\int_{t_+^*}^{t_+}\Omega(\mathbf{k},t)dt\Big|$ and $\vartheta^-(\mathbf{k})=\Big|\int_{t_-^*}^{t_-}\Omega(\mathbf{k},t)dt\Big|$, thus from Eq. (\ref{eq12}), we get a simple expression of the momentum distribution function for two pairs of turning points,
\begin{equation}
f(\mathbf{k})\approx4\cos^2[\theta(\mathbf{k})]e^{-2\vartheta(\mathbf{k})}, \label{eq17}
\end{equation}
where $\theta(\mathbf{k})\equiv\theta^{(-,+)}(\mathbf{k})=\int_{\mathrm{Re}(t_-)}
^{\mathrm{Re}(t_+)}\Omega(\mathbf{k},t)dt$. The result by \textbf{Eq. (\ref{eq17})} (dotted blue line) is plotted in Fig. \ref{Fig6} which shows the two-slit interference effect of boson longitudinal momentum spectrum in the electric field $E_{B1-}$. The oscillations of the momentum spectrum are very distinct and rapid, and the envelope value of momentum distribution function $f(k_\parallel)$ (dashed violet line) is increased to $4$ times that in single Sauter-like electric field \cite{Note1}.

From Eqs. (\ref{eq6}) and (\ref{eq7}), the numerical solution (solid red line) is calculated and compared with the approximate one. For small longitudinal momentum $k_\parallel$, the approximate solution are in good agreement with the numerical one. However, for large $k_\parallel$, the distinction between them are larger and larger. It implies that for the large longitudinal momentum the approximate solution can only qualitatively estimate the two-slit interference effect, in contrast to that for the small longitudinal momentum it can give a quantitative result. On the other hand it is worthwhile to note that, for two pairs of turning points, a more precise approximate expression has been given in Ref. \cite{Dumlu2}, which can give a consistent result with the numerical one for any longitudinal momentum \cite{Note2}.

\emph{2. N-pulse electric field}

Then, we consider the multiple-slit interference effect by using $N$-pulse electric field,
\begin{equation}
E_{B2\pm}(t)=\sum_{i=1}^N(\pm)^iE_0\mathrm{sech}^2\Big\{\omega_0 \Big[t+\Big(i-\frac{N+1}{2}\Big)T\Big]\Big\}, \label{eq18}
\end{equation}
where $E_0$, $\omega_0$ and $T$ are the same as before, "+'' and "-'' denote also the equal-sign and alternating-sign $N$-pulse electric field, respectively.

\begin{figure}[tbp]\suppressfloats
\centering
\includegraphics[width=10cm]{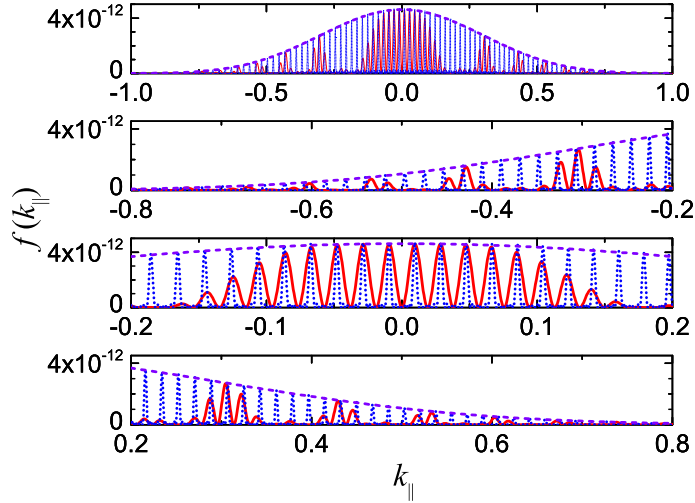}
\caption{\label{Fig7}(color online) The longitudinal momentum spectrum of created bosons in alternating-sign $N$-pulse electric field $E_{B2-}(t)$ for $N=6$. The solid red line is the numerical result, and the dotted blue line is the approximate solution. The lower three panels are the local amplification of the top one. The electric field parameters are $E_0=0.1$, $\omega_0=0.05$, and $T=180.32$.}
\end{figure}

For equal-sign electric field pulse trains $E_{B2+}(t)$, there is still one pair of complex conjugate turning points dominating the boson pair production. Therefore, no interference effect presents in the momentum spectrum of created bosons. To see it clearly, we choose the pulse number of electric field $N=6$. Its longitudinal momentum spectrum is plotted in Fig. \ref{Fig5}(b). One can see that the momentum spectrum in single pulse electric field separates from each other and no interference effect occurs. This is because the particles produced earlier can be accelerated for a longer time and attain a larger momentum than that produced later.

\begin{figure}[tbp]\suppressfloats
\centering
\includegraphics[width=10cm]{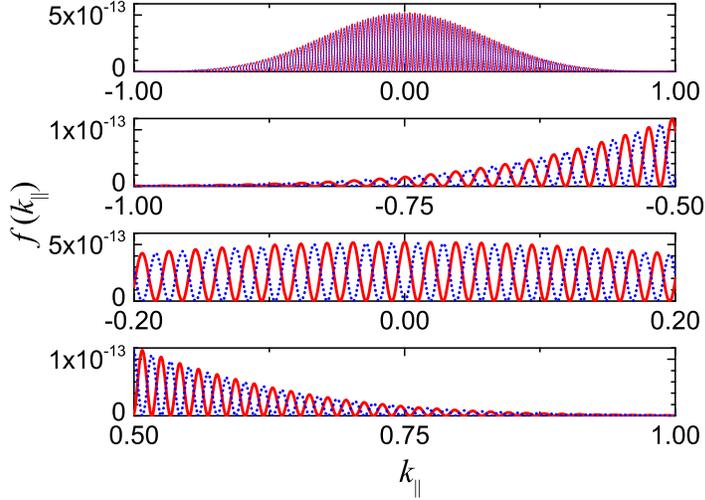}
\caption{\label{Fig8}(color online) The longitudinal momentum spectra of created bosons (dotted blue line) and fermions (solid red line) in alternating-sign $2$-pulse electric field $E_{B1-}(t)$. The lower three panels are the local amplification of the top one. The electric field parameters are $E_0=0.1$, $\omega_0=0.05$, and $T=180.32$.}
\end{figure}

For alternating-sign $N$-pulse electric field $E_{B2-}(t)$, there are $N$ complex conjugate pairs of turning points which are equidistant from the real axis and have the same distance between different pairs. Hence, the multiple-slit interference effect of boson momentum spectrum is expected. Assuming that the integral $\vartheta^{P}(\mathbf{k})$ for each pairs of tuning points is equal to $\vartheta(\mathbf{k})$ and the phase integrals $\theta^{(P,P^\prime)}(\mathbf{k})$ for different pairs of turning points are integral multiples of $\theta(\mathbf{k})$ as in Ref. \cite{Akkermans}, then we can get a simple expression of the approximate solution Eq. (\ref{eq12}) for $N$ pairs of turning points, i.e.,
\begin{equation}
f(\mathbf{k})\approx\frac{\sin^2[N\theta(\mathbf{k})]}{\sin^2[\theta(\mathbf{k})]}
e^{-2\vartheta(\mathbf{k})}. \label{eq19}
\end{equation}

The approximate solution from Eq. (\ref{eq19}) (dotted blue line) and precise numerical result from Eqs. (\ref{eq7}) and (\ref{eq8}) (solid red line) for $N=6$ are shown in Fig. \ref{Fig7}. It can be seen that the multiple-slit interference effect of boson momentum spectrum occurs and the envelope value of longitudinal momentum distribution function $f(k_\parallel)$ (dashed violet line) is increased to $6^2$ times that in single Sauter-like pulse electric field. Moreover, we find that the oscillations of momentum spectrum are very rapid and compact for approximate solution, while they are sparse for numerical result beyond the region of small longitudinal momentum. This makes clear that the consistence between the approximate solution and the numerical one is better for small than for large longitudinal momentum. The reason is that the assumption of the phase integrals for different pairs of turning points is not good enough to give a quantitative result because the time-dependent quantity $\Omega(\mathbf{k},t)$ is simply assumed as time-independent. Even so, both the approximate solution and the numerical one can give the same maximum value of momentum distribution function. Again it is about $N^2$ ($N=6$) times that in single Sauter-like pulse electric field \cite{Note1}.

To see the effect of the time delay $T$ on the momentum spectrum of created bosons, the momentum distribution function $f(k_\parallel=0)$ changing with the time delay (dotted blue line) is shown in Fig. \ref{Fig10}. We can see that there is an periodic structure with a main central peak and some side peaks in each cycle. This structure follows the standard Fabry-Perot form.

\begin{figure}[tbp]\suppressfloats
\centering
\includegraphics[width=10cm]{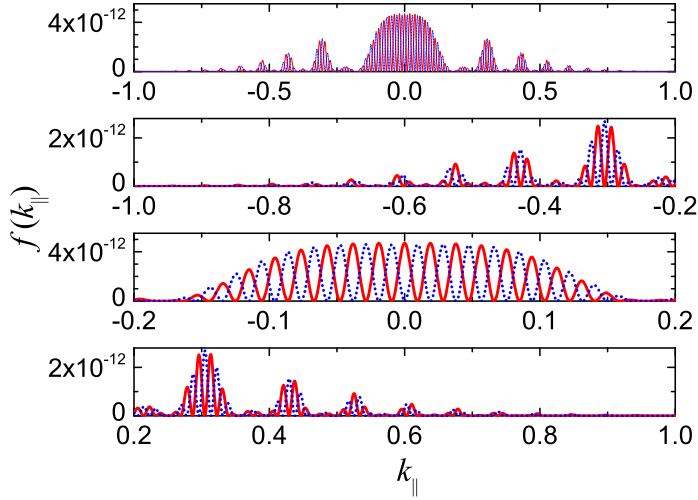}
\caption{\label{Fig9}(color online) The longitudinal momentum spectra of created bosons (dotted blue line) and fermions (solid red line) in alternating-sign $N$-pulse electric field $E_{B2-}(t)$ for $N=6$. The lower three panels are the local amplification of the top one. The electric field parameters are $E_0=0.1$, $\omega_0=0.05$, and $T=180.32$.}
\end{figure}

\subsection{Momentum spectra comparison between bosons and fermions}

Here the comparisons of momentum spectra between bosons and fermions in dynamical assist effect and multiple-slit interference effect are performed. Note that the computational formula of fermion pair production in Ref. \cite{Dumlu1} is used directly.

For dynamical assist effect, Fig. \ref{Fig1} compares the momentum spectra of created bosons (dashed blue line) and fermions (solid red line) in equal-sign electric field $E_{A1}(t)$, see Fig. \ref{Fig1}(a), and alternating-sign electric field $E_{A2}(t)$, see Fig. \ref{Fig1}(b), with the time delay $T=0$. We find that the enhancement of pair creation is weaker for bosons than for fermions, and the oscillation frequency for bosons is a little lower than that for fermions.

\begin{figure}[tbp]\suppressfloats
\centering
\includegraphics[width=10cm]{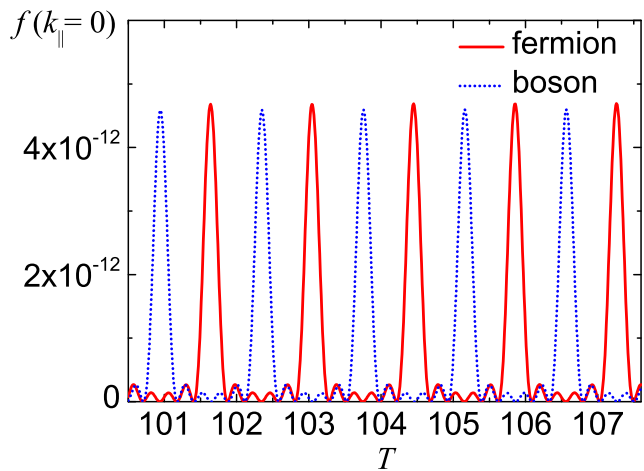}
\caption{\label{Fig10}(color online) The momentum distribution function $f(k_\parallel=0)$ changing with time delay $T$ for alternating-sign $N$-pulse electric field $E_{B2-}(t)$ for $N=6$. The solid red line is denoted as fermions and the dotted blue line is bosons. The electric field parameters are $E_0=0.1$, and $\omega_0=0.05$.}
\end{figure}

For multiple-slit interference effect, the comparisons of momentum spectra between bosons (dotted blue line) and fermions (solid red line) in alternating-sign $2$-pulse electric field $E_{B1-}(t)$ and $N(N=6)$-pulse electric field $E_{B2-}(t)$ are shown in Fig. \ref{Fig8} and Fig. \ref{Fig9}, respectively. Once again, the different oscillatory structures of momentum spectrum for bosons and fermions are presented. We can see that in a period of oscillation when the momentum distribution function $f(k_\parallel)$ for bosons increases to a maximum value, for fermions it drops down to a minimum value, and vice versa. In fact, the oscillation behaviors of momentum spectrum for bosons and fermions have a phase difference of odd multiple of $\pi/2$. This can be seen clearly from the approximate expressions of bosons (Eqs. (\ref{eq17}) and (\ref{eq19}) here) and fermions (Eqs. (6) and (13) in Ref. \cite{Akkermans}).

Moreover, from Fig. \ref{Fig10}, we show the different variations of momentum distribution function $f(k_\parallel=0)$ changing with the time delay $T$ between bosons and fermions in alternating-sign $N(N=6)$-pulse electric field. There are two distinct features presented. First, the values of $f(k_\parallel=0)$ for bosons are a little smaller than that for fermions. Second, there is also an phase difference of $f(k_\parallel=0)$ changing with the time delay $T$ between bosons and fermions.

These results reflect the obvious differences of vacuum pair production for bosons and fermions. Mathematically, the different oscillatory behaviors are decided by the second term on the right hand side of Eq. (\ref{eq12}), see Refs. \cite{Dumlu1,Dumlu2}. Physically, however, the reason is that bosons with integral spin follow Bose-Einstein statistics while fermions with half-integral spin obey Fermi-Dirac statistics, see Ref. \cite{Hebenstreit2010}.

\section{Discussions and Conclusions}

From Fig. \ref{Fig5}(b), it is interesting to see that the pair production can be enhanced by simply adding the number of Sauter-like pulse electric field. However, on the one hand, this enhancement could be very small, for example, ten pulse trains just increase one order of magnitude, on the other hand, the particles produced earlier will be greatly accelerated along the same direction by the latter pulse electric fields and escape from our studied region. Thus, it is still a challenge task to really realization although it seems experimentally feasible. By the way we would like to point out that all of our numerical results have also been checked by quantum kinetic method, e.g., by solving quantum Vlasov equation. It is found that the obtained numerical results of both methods are consistent with each other.

In summary, in this paper, based on the quantum mechanical scattering model, the dynamical assist effect and multiple-slit interference effect in electron-positron pair production from vacuum is generalized to charged scalar boson pair production. For dynamical assist effect, two kinds of combinational electric fields with different time delays, i.e., equal-sign electric field and alternating-sign electric field, are considered. It shows that the oscillation intensity of longitudinal momentum spectrum and the number density of created bosons decrease with the grow of the time delay $T$. For multiple-slit interference effect, the oscillatory structures of longitudinal momentum spectrum in the equal-sign electric field pulse trains are absent. In the alternating-sign electric field pulse trains, the approximate solution of longitudinal momentum spectrum is calculated and compared with the numerical one. Compared the approximate analytical results to direct numerical ones, especially for alternating-sign $2$-pulse electric field, they can coincide with each other well for small longitudinal momentum. For the above two effects, the differences of momentum spectra between bosons and fermions are also studied. In particular, in multiple-slit interference effect, the longitudinal momentum spectra always have a phase difference of odd multiples of $\pi/2$ between bosons and fermions so that the maxima and minima of momentum distribution function are interchanged between bosons and fermions, see Figs. \ref{Fig8} and \ref{Fig9}.

Our results are expected to deepen the understanding of vacuum pair production in the view of charged scalar bosons. They are helpful to study on the particle creation problem in particle and nuclear physics, atomic and molecular physics, condensed matter physics, and astrophysics. Furthermore, they are also useful to study on Bose-Einstein condensate, superfluidity and superconductivity.

\begin{acknowledgments}
This work was supported by the National Natural Science Foundation of China (NNSFC) under the grant No.11175023, No.11335013, and partially by the Open Fund of National Laboratory of Science and Technology on Computational Physics, IAPCM, and the Fundamental Research Funds for the Central Universities (FRFCU). Numerical calculation was carried out at the HSCC of the Beijing Normal University.
\end{acknowledgments}

\end{document}